\documentclass{article}

\usepackage{spconf}
\usepackage{hyperref}       %
\usepackage{url}            %
\usepackage{booktabs}       %
\usepackage{nicefrac}       %
\usepackage{microtype}      %
\usepackage{amsfonts}       %
\usepackage{amsmath}
\usepackage{tabularx}
\usepackage{float}
\usepackage{outlines}
\usepackage{subcaption}
\usepackage{algorithm}
\usepackage{algpseudocode}
\usepackage{chngcntr}
\usepackage{graphicx}
\usepackage{xcolor}
\usepackage{setspace}
\usepackage{enumitem}
\usepackage{bbm}
\usepackage{pgfplots}%
\pgfplotsset{compat=1.13}%
\usepackage[htt]{hyphenat}
\usepackage{lipsum}
\usepackage{xspace}
\usepackage{multirow}
\usepackage[framemethod=TikZ]{mdframed}
\usepackage{tabu}
\usepackage{pifont}%

\usepackage[square,sort,comma,numbers]{natbib}
\usepackage[subtle]{savetrees}

\hypersetup{
    colorlinks,
    linkcolor={red!50!black},
    citecolor={blue!80!black},
    urlcolor={blue!80!black}
}

\renewcommand{\paragraph}[1]{\noindent\textbf{#1}\quad}

\newcommand{\cmark}{\ding{51}}%

\newcommand{\method}{CLC\xspace}
\newcommand{\methodlong}{\underline{C}ontrastive \underline{L}earning for \underline{C}onversations\xspace}
\newcommand{\dataset}{OD3\xspace}
\newcommand{\datasetlong}{\underline{O}pen \underline{D}irected \underline{D}ialogue \underline{D}ataset\xspace}

\DeclareMathSizes{10}{9}{6}{5}

\setlist{nosep} %
\setlist{itemsep=1pt, topsep=3pt}

\title{Task Oriented Dialogue as a Catalyst for Self-Supervised Automatic \\ Speech Recognition}
\name{David M. Chan$^{\star \dagger}$ \qquad Shalini Ghosh$^{\dagger}$ \qquad Hitesh Tulsiani$^{\dagger}$ \qquad Ariya Rastrow$^{\dagger}$ \qquad Bj{\"o}rn Hoffmeister$^{\dagger}$}
\address{$^{\star}$ University of California, Berkeley $\quad$
     	 $^{\dagger}$ Amazon Alexa AI}
\begin{document}
\ninept
\maketitle
\begin{abstract}
While word error rates of automatic speech recognition (ASR) systems have consistently fallen, natural language understanding (NLU) applications built on top of ASR systems still attribute significant numbers of failures to low-quality speech recognition results. Existing assistant systems collect large numbers of these unsuccessful interactions, but these systems usually fail to learn from these interactions, even in an offline fashion. In this work, we introduce \method: \methodlong, a family of methods for contrastive fine-tuning of models in a self-supervised fashion, making use of easily detectable artifacts in unsuccessful conversations with assistants. We demonstrate that our \method family of approaches can improve the performance of ASR models on \dataset, a new public large-scale semi-synthetic meta-dataset of audio task-oriented dialogues, by up to 19.2\%. These gains transfer to real-world systems as well, where we show that CLC can help to improve performance by up to 6.7\% over baselines.\footnote{Our Code/Data is publicly available at \url{https://github.com/amazon-science/amazon-od3}.}
\end{abstract}

\section{Introduction \& Background}

When users interact with assistant systems in task oriented ways, they build rich conversational contexts, which contain information that may be relevant to future requests along with feedback on the performance of the system. When users are dissatisfied, they express that intent in many ways, from direct corrections of the system response, to repeating and rephrasing the original question \cite{kwan2023survey}. This discourse provides a source of contextual user interaction signals that are relatively untapped in Automatic Speech Recognition (ASR).

Indeed, traditional systems for ASR have primarily focused on single-utterances \cite{radford2023robust,baevski2020wav2vec,hsu2021hubert,chan2022multi,chan2023domain,mitra:2023:unified}, which, although flexible, overlook the wealth of contextual cues available in task directed dialogues. While work has been done in natural language understanding (NLU) to exploit these cues \cite{min2021recent}, their potential in ASR has remained largely unexplored, primarily due to the limited availability of task-driven dialogue datasets in the audio domain \cite{chang2023context,si2023spokenwoz}. Current efforts to integrate context from non-dialogue sources into ASR often involve training models explicitly with external per-turn contextual inputs, often leveraging context attention mechanisms~\cite{chang2023context,kim2018dialog, chan2023domain,chang2021context,chen2019joint,sathyendra2022contextual,wei2021attentive,yang:2023:generative,chan:2023:off-policy,mahadevan:ecml:2018}. While per-turn context is important for the ASR task, these methods do not draw from dialogue structures, nor do they account for interactive feedback present in labeled dialogues. 
 
\begin{figure}
    \centering
    \includegraphics[width=0.7\linewidth,keepaspectratio]{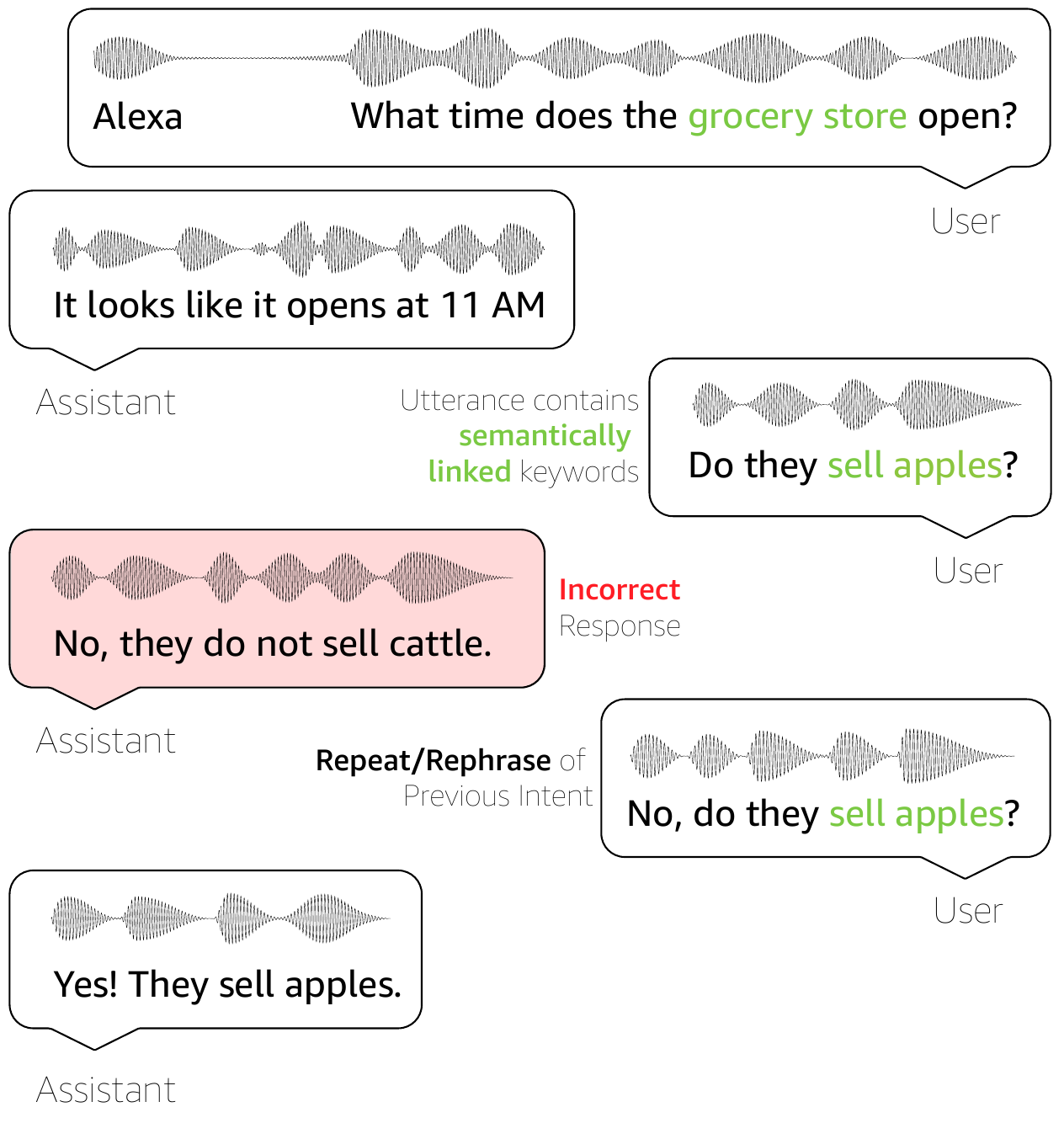}
    \caption{Task oriented dialogues can contain a multitude of relevant information for performing automated speech recognition. In this work, we explore how we can learn from both semantically linked keywords within dialogues, and failed dialogue turns.}
    \label{fig:teaser}
    \vspace{-1em}

\end{figure}

Instead of directly training on per-sample or per-turn context (e.g., contact names \cite{chen2019joint, sathyendra2022contextual}, or external dictionaries \cite{chan2023domain}), we explore the potential of learning implicit contextual signals of user interactions, which remains relatively untapped in ASR-based dialog systems. Following work demonstrating the benefits of contrastive learning in ASR \cite{chan:interspeech:2022}, closest to our work may be Chang \textit{et al.} \cite{chang2023context} who propose reducing ASR errors with contrastive learning between noisy and clean audio transcripts from task-oriented dialogues -- however, their work focuses only on single turns of dialogues, not contextual dialogue cues. Our primary contributions are:
\begin{itemize}
    \item We propose a new family of self-supervised fine-tuning losses, \method, which incorporate self-supervised information from task oriented dialogues (TODs), and show that learning from TODs, even those with errors, provides benefits over fine-tuning. 
    \item We introduce a new semi-synthetic benchmark meta-dataset, the \datasetlong (\dataset), designed to enable further research in conversational interactions for ASR.
\end{itemize}
\vspace{-1em}
\section{Contrastive Learning for Conversations}
\label{sec:clc}
\label{sec:methods}

In this work, we introduce two novel auxiliary losses, termed ``Contrastive Learning for Conversations'' (CLC),  designed to enable learning from both successful and unsuccessful task-directed conversations with assistants (\autoref{sec:clc}), as well as a new synthetic dataset for the evaluation of contextual automated speech recognition models in task directed domains (\autoref{sec:dataset}).

\paragraph{Learning from Past and Future Dialogues:}  As shown in \autoref{fig:teaser}, utterances in a dialogue can contain important contextual hints useful for recognizing low-frequency words in the sentence. While we may not have access to past or future utterances at inference, we can often learn from these hints during training. The first auxiliary loss we introduce follows this key motivation; auditory information within a dialogue should share more semantic and representational overlap than auditory information from a second, unrelated dialogue. 

This insight induces a natural contrastive loss: the speech encoder representations of audio within a session should be closer in the latent space (on average) than the representations between sessions. To implement a ``Past-Future" contrastive loss, we consider the utterances $u_1,\dots,u_N$ in a dialogue. Let the speech encoder be defined as $e_i = \epsilon(u_i) \in \mathbb{R}^{T \times k}$, where $k$ is the dimension of the speech encoder embedding, and T is the number of frames of audio in the dialogue. We further introduce three ``head" encoders, $\xi_{past}(e_i) \in \mathbb{R}^d, \xi_{current}(e_i) \in \mathbb{R}^d, \xi_{future}(e_i) \in \mathbb{R}^d$, which embed the sequential embeddings from the encoder $\epsilon$ of the current, past, and future frames into single vectors (of dimension $d$) representing the current, past, and future contexts. These head encoders take the form of global pooling followed by two layers of a shallow MLP with ReLU activations, LayerNorm, and Dropout. We can then compute the following contrastive loss terms (similar to \cite{khosla2021supervised}) for a batch of $1 \le i,j \le N$ samples (where embeddings are L2-normalized):

\begin{figure}
    \centering
    \includegraphics[width=0.9\linewidth]{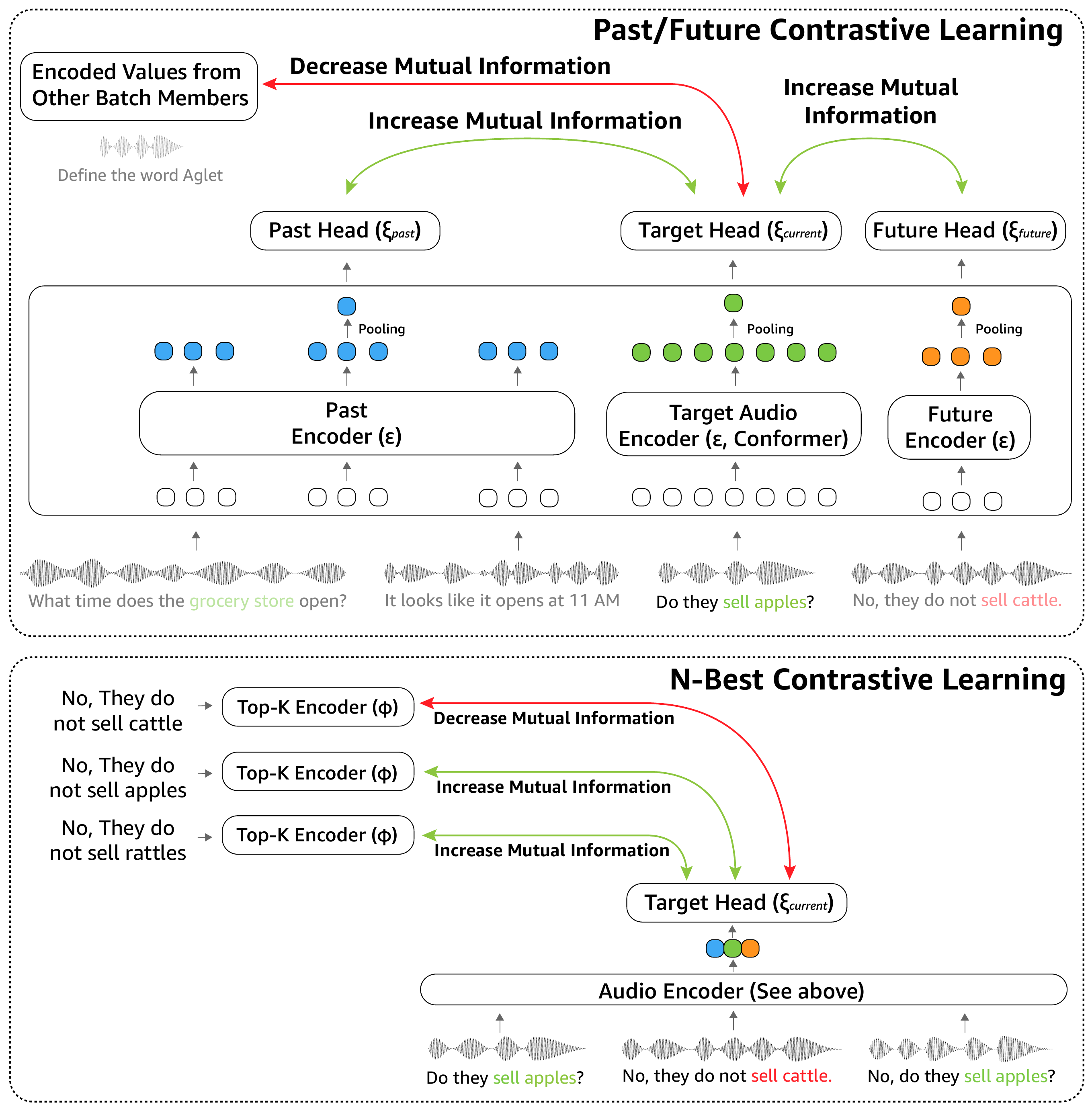}
    \caption{Overview of \method approaches. The Past-Future loss maximizes agreement between current, past, and future embeddings. The N-best loss minimizes agreement between current embeddings and top predictions of rephrases, while maximizing agreement otherwise.}
    \label{fig:methods}
    \vspace{-1em}
\end{figure}

\begin{align}
L_{future}^{i,j} & = -\log\left[\frac{\exp(\xi_{current}(e_i) \cdot \xi_{future}(e_j)/\tau)}{\sum_{k=1}^{N}\exp(\xi_{current}(e_i) \cdot \xi_{future}(e_k)/\tau)}\right] \nonumber \\
L_{past}^{i,j} & = -\log\left[\frac{\exp(\xi_{current}(e_i) \cdot \xi_{past}(e_j)/\tau)}{\sum_{k=1}^{N} \exp(\xi_{current}(e_i) \cdot \xi_{past}(e_k)/\tau)}\right] \nonumber
\end{align}

The ``Past-Future'' auxiliary loss is then a weighted sum:

\begin{equation}
    L_{pf} = \frac{1}{N}\left[\alpha\sum_{i=1}^NL_{future}^{i,i} + \beta\sum_{i=1}^NL_{past}^{i,i} 
    \right] 
\end{equation}

Here, we choose cosine distance (the dot-product) as the similarity function. The result of the loss function is that we aim to maximize the mutual information between the encoding of the current, future and past frames within a dialogue, while minimizing the mutual information between the current frames and frames from other dialogues. Note here that it’s important that $e_i \neq e_j$, that is, the embedding of the past should not be identical to the embedding of the future (as they have different ASR content). Instead, we encourage high mutual information between the different segments, by leveraging contrastive learning on projection heads stemming from the shared representation. $\alpha$ and $\beta$ are hyper-parameters which control the strength of the binding in the loss function, and $\tau$ is a temperature parameter. In our experiments, we found through grid hyper-parameter search of $\alpha, \beta \in [0.0001, 100]$ (logarithmic sweep) and $\tau \in [0.1, 1]$ (linear sweep) that $\alpha=1.0, \beta=0.7, \tau=0.1$ is the most effective.

\paragraph{Learning from Failures:} We can extract valuable semantic information from conversations, even those that don't proceed smoothly. It is often possible to detect dialogues where unsuccessful ASR has triggered repeats and rephrases of previous content by understanding when subsequent user turns have high semantic overlap, or tracking NLU failures in downstream systems. In these cases, we can further leverage contrastive learning to improve the performance of the model. Ideally, when there is a repeat or rephrase in a dialogue, we want to reduce the mutual information between the conformer encoder embedding of the initial turn triggering the repeat or rephrase, and the answer produced by the model in that dialogue. While we could use reinforcement learning to optimize for this signal (and it is interesting future work to do so), we often train models offline, and as the model trains, its decisions deviate from the original policy, leading to a breakdown in the learning process. Instead, as we know the ``bad" solution, we can use supervised contrastive learning \cite{khosla2021supervised} to improve the model. When there is no rephrase, we want to increase the mutual information between the semantics of the top-1 prediction of the model and the current frames. When there is a rephrase, we want to decrease the mutual information between the semantics of the top-1 prediction, and instead encourage the model to produce a different output from the top-k. While it is possible that worse hypotheses with high similarity exist in the hypothesis set (leading to incorrect labels), we observe empirically that our models have high oracle WER, allowing this method to achieve a weak approximation to oracle re-ranking of the candidate set, which improves overall performance when smoothed over a large training set. 

An overview of our n-best approach is given in \autoref{fig:methods}. For each sample $u_i$, let $\phi_{1}(u_i) \dots \phi_{K}(u_i)$ be the semantic embeddings of the top-k predictions of the i’th utterance (using beam-search decoding) and $\xi_{current}(e_i)$ be an embedding of $e_i$ for $u_i$.  Using a similar set of heads to the network above, we compute positive and negative losses:
\begin{equation}
    L_{pos}^i = -log\left[\frac{\exp(\xi_{current}(e_i) \cdot \phi_1(u_i)/\tau) }{\sum_{k=1}^K\exp(\xi_{current}(e_i) \cdot \phi_k(u_j)/ \tau) }\right] \nonumber
\end{equation}
\begin{equation}
    L_{neg}^{i} = -log\left[\frac{\max_{j\neq i}\left[\exp(\xi_{current}(e_i) \cdot \phi_j(u_i) / \tau)\right]}{\sum_{k=1}^K\exp(\xi_{current}(e_i) \cdot \phi_k(u_i) /\tau) }\right] \nonumber
\end{equation}
Let $\mathcal{R}$ be the set of utterances which trigger a repeat/rephrase, and $\mathcal{S}$ be the set of utterances which are considered successful. We can then combine the positive and negative losses as follows:
\begin{align}
\begin{split}
    L_{nbest}  = &  \frac{\gamma}{|\mathcal{R}|}\sum_{i \in \mathcal{R}}  L_{neg}^i + \frac{\kappa}{|\mathcal{S}|} \sum_{i \in \mathcal{S}} L_{pos}^i z
\end{split}
\end{align}

\noindent where $\gamma$ and $\kappa$ are hyper-parameters controlling the trade-off between negative and positive reinforcement. Discovering the sets $\mathcal{S}$ and $\mathcal{R}$ can be challenging, however, we can detect repeats and rephrases with relatively high accuracy using semantic vector matching (such as matching BERT embeddings). Using grid search with $\gamma,\kappa \in [0.0001, 100]$ (logarithmic sweep), we found $\gamma=0.1,\kappa=1.0$ was most effective.

\vspace{-0.5em}
\subsection{Data}
\label{sec:dataset}

While a predominant portion of interactions with assistant systems revolves around task-directed dialogues, the availability of datasets (\autoref{tab:odds}) encompassing task-directed audio interactions remains quite limited. Moreover, even within datasets that do incorporate such interactions, a conscious effort has been exerted to remove flawed turns (turns in which a dialogue assistant responds incorrectly, and must be corrected by the user). To evaluate our \method methods for self-supervised fine-tuning, we use two datasets: a private collection of de-identified real-world conversations with a conversational assistant, and a new semi-synthetic meta-dataset, \dataset, replicating flawed conversations often seen in real-world assistant interactions. The \dataset dataset is released as part of this work under the CC-BY-NC-SA (4.0) license.

\vspace{-0.5em}
\subsubsection{Real-World (Internal) Data}

To demonstrate the performance of our method, we train and evaluate our models on 130K hours of de-identified agent-centric task-directed dialogues constructed from independent interactions with a conversational assistant. These dialogues have a maximum of five utterances each (with an average of 1.2 turns per goal). Dialogues are constructed around a seed utterance by collecting interactions within $\rho=90$ seconds on each size of the utterance. This process is repeated recursively until there are no more interactions. In the case that there are more than five utterances, we halve $\rho$, and repeat the process. This continues until either we have less than 5 utterances in the final set or we hit a minimum time gap of $15$ seconds.  During testing, only the past and current context is available to the model (the future remains hidden).

\vspace{-0.5em}
\subsubsection{\dataset: A new dataset for conversational learning} 

In addition to the results on real-world interactions in this paper, we further introduce a new semi-synthetic meta-dataset, \dataset (\datasetlong), which is designed to allow the community to explore further research into leveraging flawed conversational interactions to improve model performance. \dataset is a collection of 63K conversations (600K turns, 1,172 hours of audio) drawn from existing natural language task-oriented dialog datasets, and augmented with synthetic audio. \dataset is further augmented with turns containing \textit{repeats} and \textit{rephrases} of previous failed utterances. We compare our dataset with some others in the field in \autoref{tab:datastats}.

\paragraph{Constructing \dataset:} To construct \dataset, we start with several seed datasets of natural language task oriented dialog data: KVRET \cite{eric2017key}, Multi-Woz \cite{budzianowski2018multiwoz}, DSTC11 (Track 5) \cite{zhao2023what}, NOESIS-II \cite{gunasekara2020noesis} and SIMMC-2.1 \cite{kottur2021simmc}. Here we focus on multi-turn dialogue, (as opposed to single-turn datasets such as those for question-answering like NMSQA \cite{lin2022dual}), as they contain the most relevant contextual information. This gives us a pool of $\approx$ 63K unique dialogues ($\approx$ 597K turns) containing no explicitly labeled errors or flaws. Because these datasets are not augmented with audio for each of the conversational turns, we leverage the NeMo Text Normalizer \cite{kuchaiev2019nemo} and the YourTTS method \cite{casanova2022yourtts} (voice cloning) to generate audio for each of the conversations. In all of the conversations, we hold the voice for the agent constant, and each voice used in voice cloning is randomly selected from the English subset of Common Voice \cite{ardila2019common} (which is CC0 licensed). We found that in some cases, the TTS induces errors in the generated speech, which we found correlated with a high number of deletions in the resulting ASR models. To clean the dataset, we filter out $\approx$4K utterances inducing a significant number of deletions in both our tested and third party ASR models. While we run our experiments in this paper on the clean data, we additionally release the noisy versions of the data as they could be useful for investigation into alternate directions of research.

We synthetically introduce errors and noisy conversations into the data. For that, we first compute ASR for each dialog turn using OpenAI's Whisper Large (v2) model \cite{radford2023robust}. We consider conversational turns with WER higher than 15\% candidates for the injection of either a \textit{repeat}, or a \textit{rephrase} of the intent. We then insert repeats and rephrases into 20\% of the possible candidate conversations. To insert a \textit{repeat}, we introduce two conversational turns: a response for the agent which is a non-specific error response (such as ``I'm sorry, I don't understand"), and a repeat of the phrase which triggered the ASR errors (re-sampled from the original TTS model). Inserting a \textit{rephrase}, on the other hand, is much more complicated. Similar to the case of repeats, we first introduce a non-specific agent error message. We then generate a rephrase of the original triggering utterance using in-context learning with the MPT-30B language model \cite{MosaicML2023Introducing}, combined with the prompt: \texttt{Our automated speech recognition model\- found "<input string>" hard to parse, so we rephrased it to use easier to understand words as "...}

We found that this prompt generated reasonable rephrases of the candidate sentences. For example, \textit{``Are there noisy neighbors?"} was rephrased as \textit{``Is the place quiet enough?"}.  This gives us a total of $\approx$ 625K turns of dialogue in $\approx$ 62K sessions, and 1,172 hours of audio.

\begin{table}[t]
    \caption{Statistics for \dataset. \dataset is much larger than existing TOD datasets, while including both audio and noisy conversations.}
    \footnotesize
    \centering
    \begin{tabularx}{\linewidth}{Xccccc}
    \toprule
    \textbf{Dataset} & \textbf{Dialogues} & \textbf{Turns} & \textbf{Audio} & \textbf{Errors} \\
    \midrule
    {DSTC-2 \cite{henderson2014second}} & 1,612 & 23,354 & \cmark & \\
    {KVRET \cite{eric2017key}} & 2,425 & 12,732  & & \\
    {MultiWOZ \cite{budzianowski2018multiwoz}} & 8,438 & 115,424  & & \\
    {DSTC-10 \cite{kim2021robust}} & 107 & 2,292 & & \\
    {SpokenWOZ \cite{si2023spokenwoz}} & 5,700 & 203,074 & \cmark &  \\
    \textbf{\dataset (Ours)}  & \textbf{62,974} & \textbf{623,145} & \cmark & \cmark \\
     \bottomrule
    \end{tabularx}
    \label{tab:datastats}
\end{table}

\subsection{Models}

For the speech encoder $\epsilon$, we use a conformer architecture \cite{gulati2020conformer}, with 17 layers, latent dimension of 1024, and two stride-two convolutional sub-sampling layers  ($\approx$ 200M parameters). We use a 1-layer LSTM decoder with latent dimension of 320, and a 4K token vocabulary. The encoder/decoder are initialized with a model pre-trained on 120K hours of de-identified internal seed data.  During training, we apply both kernel regularization and bias regularization with weight $1e^{-6}$, and dropout with weight $1.0$. 
We optimize the overall loss:
\begin{equation}
    L_{overall} = L_{asr} + \lambda L_{pf} + \delta L_{nbest}
\end{equation}
The models are trained for at most 120 epochs with the Adam optimizer, following a linear increase, hold, exponential decay learning rate schedule starting at $1e^{-8}$, increasing linearly over 50K steps to hold at $4e^{-5}$ for 250K steps, and then decay back to $1e^{-6}$ over a further 300K steps. 
We use gradient clipping with limit $0.3$, and a dynamic batch size (depending on input feature length) ranging between 128 and 1024. 
As contrastive learning cannot naively be scaled across GPUs, we leverage techniques similar to BASIC \cite{pham2023combined} and perform memory efficient contrastive mini-batching.

\section{Results \& Discussion}
\label{sec:results}

\begin{table}
\scriptsize
\caption{\small Results on internal data, both overall and only on turns inducing repeats or rephrases. WERR ($\uparrow$): Percent relative WER Improvement. SERR ($\uparrow$): Percent relative SER improvement.}\label{tab:alexa}
\begin{tabularx}{\linewidth}{Xcccc}
	\toprule
                            \multirow{2}{*}{\textbf{Model}} & \multicolumn{2}{c}{\textbf{Overall}} & \multicolumn{2}{c}{\textbf{Repeats/Rephrase}} \\
		& \textbf{WERR} & \textbf{SERR} & \textbf{WERR} & \textbf{SERR}  \\
	\midrule
        Zero-Shot (No Fine Tuning) & -23.02\% & -17.46\% & -4.65\% & -5.75\% \\
        Baseline (Fine Tuned) & - & -  & - & - \\
        \midrule
        \method ($\lambda=1,\delta=0$) & 2.75\% & 2.88\% & 3.0\% & 3.39\% \\
        \method ($\lambda=0,\delta=1$) & 2.60\% & 2.39\% & 3.75\% & 3.87\% \\
        \method ($\lambda=1,\delta=1$) & 4.31\% & 3.88\% & 5.07\% & 5.31\% \\
	\bottomrule
\end{tabularx}
\end{table}

\begin{table}
\scriptsize
\caption{\small Results on internal data for different values of $\alpha$ and $\beta$ ($\tau=0.1$) in $L_{pf}$,  as well as $\gamma$ and $\kappa$ in $L_{nbest}$ for small scale (batch size 128) experiments. WERR ($\uparrow$): Relative WER Improvement. SERR ($\uparrow$): Relative SER improvement.}\label{tab:alexa_alpha}
\begin{tabularx}{\linewidth}{Xcc}
	\toprule
		\textbf{Model} & \textbf{WERR} & \textbf{SERR} \\
	\midrule
        Baseline (\method, $\lambda=0,\delta=0$) & - & - \\
        \midrule
        \method ($\alpha=1,\beta=0$\textcolor{gray!40}{,$\gamma=0$, $\kappa=0$}) & \textcolor{black}{3.28\%} & \textcolor{black}{2.26\%} \\
        \method ($\alpha=0,\beta=1$\textcolor{gray!40}{,$\gamma=0$, $\kappa=0$}) & \textcolor{black}{2.74\%} & \textcolor{black}{3.68\%} \\
        \method ($\alpha=1,\beta=1$\textcolor{gray!40}{,$\gamma=0$, $\kappa=0$}) & \textcolor{black}{4.50\%} & \textcolor{black}{5.34\%} \\
        \method ($\alpha=1,\beta=0.7$\textcolor{gray!40}{,$\gamma=0$, $\kappa=0$}) & \textcolor{black}{5.17\%} & \textcolor{black}{4.67\%} \\
        \midrule
        \method (\textcolor{gray!40}{$\alpha=0,\beta=0$,}$\gamma=1,\kappa=0$) & \textcolor{black}{-11.81\%} & \textcolor{black}{-10.43\%} \\
        \method (\textcolor{gray!40}{$\alpha=0,\beta=0$,}$\gamma=1,\kappa=1$) & \textcolor{black}{-1.88\%} & \textcolor{black}{-2.21\%} \\
        \method \textcolor{gray!40}{$\alpha=0,\beta=0$,}$\gamma=0,\kappa=1$) & \textcolor{black}{6.23\%} & \textcolor{black}{5.59\%} \\
        \method (\textcolor{gray!40}{$\alpha=0,\beta=0$,}$\gamma=0.1,\kappa=1.0$) & \textcolor{black}{6.77\%} & \textcolor{black}{6.25\%} \\
	\bottomrule
\end{tabularx}
\end{table}

We first demonstrate the performance of our method on our internal session data. From the results in \autoref{tab:alexa}, we can see that all three settings of \method improve the overall WER/SER of the model, particularly over zero-shot models. We notice that setting $\lambda=1$ is the most effective at reducing overall WER, as in most situations, contextual information from previous (and future) turns can provide more powerful hints to the content of an utterance. While $\delta$ is helpful as well, it is less important to overall WER. 

\autoref{tab:alexa_alpha} shows the performance of \method across different values of $\alpha$ and $\beta$ for $L_{pf}$. We can see that taking into account both past and future information is important. Unsurprisingly, past information is a more powerful indicator of the current ASR context; however it's important to note that pre-training with the information from the future allows the model to improve the predictive ability of the audio representations, leading to improvements (particularly in SER). \autoref{tab:alexa_alpha} also shows the performance for  values of $\gamma$ and $\kappa$ in the $L_{nbest}$ loss. We can see here that placing too much weight on the $\gamma$ term leads to a destabilization of the loss, however small magnitude $\gamma$ values can help with overall performance. We believe that this destabilization is caused by the high variance of the $\max_{j\neq i}\left[\exp(\xi_{current}(e_i) \cdot \phi_j(u_i) / \tau)\right]$ term, and it is future work to explore how functional implementations such as a soft-max could reduce the gradient variance stemming from this loss term. 

\autoref{tab:alexa} also shows the performance of our method when restricted to only defective utterances: utterances triggering repeats and rephrases in the dataset. We can see that setting $\delta=1$ is helpful, since the additional losses nudge the model away from high-confidence decisions in detected repeats/rephrases and makes an impact on the model's ability to correctly recognize challenging samples. Note that WERR/SERR gains are statistically significant over the large-scale test set ($\approx 1K$ hours of test audio). 

\begin{table}
\scriptsize
\caption{\small Results on the \dataset dataset (overall and repeat/rephrase inducing). WER ($\downarrow$): Word Error Rate, BERT-S ($\uparrow$): Bert Score.}\label{tab:odds}
\begin{tabularx}{\linewidth}{Xcccc}
	\toprule
            \multirow{2}{*}{\textbf{Model}} & \multicolumn{2}{c}{\textbf{Overall}} & \multicolumn{2}{c}{\textbf{Repeat/Rephrases}} \\
		& \textbf{WER} & \textbf{BERT-S} & \textbf{WER} & \textbf{BERT-S} \\
        \midrule
        Baseline (206M) & 11.13 & 0.9762 & 16.17 & 0.9690 \\
        \midrule
        \method ($\lambda=1,\delta=0$) & 9.57 & 0.9801 & 14.12 & 0.9702 \\
        \method ($\lambda=0,\delta=1$) & 9.38 & 0.9803 & 13.94 & 0.9721 \\
        \method ($\lambda=1,\delta=1$) & 8.99 & 0.9812 & 13.81 & 0.9737 \\
	\bottomrule
\end{tabularx}
\end{table}

\begin{table}
\scriptsize
\caption{\small Zero-shot results on \dataset for several open-source models. Models in this table are not directly comparable (trained on differing data, setups, hyperparameters, optimizers etc.), but serve as a benchmark for performance on \dataset under several varying setups. WER ($\downarrow$): Word Error Rate, BERT-S ($\uparrow$): Bert Score.}\label{tab:external}
\begin{tabularx}{\linewidth}{Xcccc}
	\toprule
             \multirow{2}{*}{\textbf{Model}} & \multicolumn{2}{c}{\textbf{Overall}} & \multicolumn{2}{c}{\textbf{Repeat/Rephrases}} \\
		 & \textbf{WER} & \textbf{BERT-S} & \textbf{WER} & \textbf{BERT-S} \\
	\midrule
        \method best model & 8.99 & 0.9812 & 13.81 & 0.9737 \\
        Whisper S (200M) \cite{radford2023robust} & 11.24 & 0.9775 & 14.17 & 0.9727 \\
        Whisper L (1.3B) \cite{radford2023robust} & 8.51 & 0.9852 & 12.37 & 0.9792   \\
        Conformer (100M, Librispeech) \cite{gulati2020conformer} & 19.26 & 0.9612 & 22.19 & 0.9571\\
        Wav2Vec 2 (433M, Librispeech) \cite{baevski2020wav2vec}  & 19.41 & 0.9582 & 22.03  & 0.9544 \\
        Streaming Conformer (45M) \cite{tsunoo2021streaming}  & 14.38 & 0.9701 & 16.70 & 0.9665 \\
	\bottomrule
\end{tabularx}
\end{table}

On \dataset, our approach produces even more defined results, demonstrated in \autoref{tab:odds}, where our model produces a 19.22\% improvement over baselines, clearly showing how learning from additional contextual clues can benefit ASR models. Interestingly, despite a high word error rate, the semantic similarity, as indicated by the BERTScore \cite{zhang2019bertscore} remains high --- this suggests that ASR errors, while numerous, do not significantly impact the semantic meaning. Several major questions remain unanswered for future work, for example, it remains an open question how the approaches scale with model parameters, as well as understanding to what extent different mixes of pre-training data alter the performance of the model. 

Even for models with strong language models, large vocabularies, and training data focused on open-domain conversational language, \autoref{tab:external} shows that \dataset is challenging. Models demonstrated increased insertions and substitutions, as there are a large number of challenging low-frequency words that must be recognized accurately. It's interesting to see that the streaming conformer \cite{tsunoo2021streaming} (trained on Gigaspeech) outperforms some of the larger models. This is likely due to the training data mix: training smaller models on more robust datasets is more effective than training larger models on sparse or biased data.  
\vspace{-0.5em}
\section{Conclusion}

This work introduces \method, a self-supervised fine-tuning approach for enhancing contextual automated speech recognition (ASR) in task-oriented dialog systems. We also introduced \dataset, the largest-ever dataset for task-oriented automated speech recognition. 
By leveraging both successful and unsuccessful conversational interactions, our method enhances the underlying ASR model's ability to handle challenging and contextually rich utterances. In real-world data, we demonstrate as much as 6.77\% improvement over baselines. Further, for \dataset we show up to a 19.22\% improvement over baselines. 
We hope that our approaches and datasets will help address ASR challenges within intricate and error-prone dialog settings, elevating user experiences and enabling more effective interactions between humans and AI agents.
\clearpage
\section{References}
\begingroup
  \def\section*#1{}
  \small
  \setlength{\bibsep}{4pt}
  \bibliographystyle{IEEEtranN}
  \bibliography{refs}
\endgroup

\end{document}